\documentclass[hyphens]{article}

\PassOptionsToPackage{numbers, compress}{natbib}
 \usepackage[position,preprint]{neurips_2025}

\usepackage[utf8]{inputenc} 
\usepackage[T1]{fontenc}    
\usepackage[hyphens]{url}    
\usepackage{hyperref}       
\usepackage{booktabs}       
\usepackage{amsfonts}       
\usepackage{nicefrac}       
\usepackage{microtype}      
\usepackage{xcolor}         

\usepackage{todonotes}
\usepackage{cleveref}
\usepackage{tabularx}
\usepackage{multirow} 

\title{Tiered Anonymity on Social-Media Platforms as a Countermeasure against Deepfakes and LLM-Driven Mass Misinformation}

\author{%
$\textbf{David Khachaturov}^{\dagger*}$ \quad $\textbf{Roxanne Schnyder}^{\ddagger*}$ \quad $\textbf{Robert Mullins}^{\dagger}$\\
$^\dagger$Department of Computer Science and Technology \quad $^\ddagger$Institute of Criminology \\
University of Cambridge
}

\begin{document}

\maketitle

{\def\thefootnote{*}\footnotetext{Equal contribution}}

\begin{abstract}
We argue that governments should mandate a three-tier anonymity framework on social-media platforms as a reactionary measure prompted by the ease-of-production of deepfakes and large-language-model-driven misinformation. The tiers are determined by a given user's \textit{reach score}: Tier 1 permits full pseudonymity for smaller accounts, preserving everyday privacy; Tier 2 requires private legal-identity linkage for accounts with some influence, reinstating real-world accountability at moderate reach; Tier 3 would require per-post, independent, ML-assisted fact-checking, review for accounts that would traditionally be classed as sources-of-mass-information.

An analysis of Reddit shows volunteer moderators converge on comparable gates as audience size increases -- karma thresholds, approval queues, and identity proofs -- demonstrating operational feasibility and social legitimacy. Acknowledging that existing engagement incentives deter voluntary adoption, we outline a regulatory pathway that adapts existing US jurisprudence and recent EU-UK safety statutes to embed reach-proportional identity checks into existing platform tooling, thereby curbing large-scale misinformation while preserving everyday privacy.
\end{abstract}

\section{Introduction}

\textbf{Governments should mandate \emph{tiered anonymity} on social-media platforms to curb the democratic harms of deepfakes and large-language-model–driven misinformation.} When influence is algorithmically amplified and truth is algorithmically optional, the notion that all online voices should enjoy equal anonymity becomes not a right, but a liability. This position responds to the growing asymmetry between the ease with which synthetic content can shape public discourse and the absence of mechanisms to hold the most influential voices accountable.  Generative models now enable anyone to manufacture persuasive audio-visual fabrications at negligible cost, eroding the traditional evidentiary value of sight and sound and fueling the ``liar's dividend'', the tactic of dismissing inconvenient truths as fakes~\cite{Chesney_Citron_2019,liars_divident_2024}. Simultaneously, recommender systems amplify attention without regard to veracity, allowing fringe messages to reach millions in minutes.

Online anonymity began as one of the internet’s core protective mechanisms, allowing ordinary people, dissidents, and vulnerable groups to speak freely without fear of retaliation. However, when algorithmic amplification gives a single post the reach of a broadcaster, blanket anonymity becomes a public-safety liability. We therefore argue that identity obligations should scale with influence. Crucially, the aim is not to censor speech but to regulate its amplification and reach, preserving free expression while curbing the outsized impact of unverified high-reach accounts.

Our proposed three-tier model, summarized in~\Cref{table:tiers_summary}, assigns obligations by some notion of \emph{reach} (e.g.\ a weighted sum of followers, shares, views, etc.). Tier 1 preserves full pseudonymity\footnote{We use ``pseudonymity'' to mean the use of persistent screen names not linked to offline identity} for low-reach accounts; Tier 2 requires a platform-held legal-identity link once a predefined reach threshold is crossed; Tier 3 adds independent, ML-assisted fact-checking for mass-reach content.

\begin{table}[t]
\centering
\caption{Proposed tiered anonymity framework. Tier thresholds are discussed in~\Cref{sec:high_level_implementation}, are generally illustrative, and should be calibrated per-platform. A single post that crosses a threshold retroactively elevates the account to the corresponding tier.}
\label{table:tiers_summary}
\begin{tabularx}{\textwidth}{@{}c l X@{}}
\toprule
\textbf{Tier} & \textbf{Typical Accounts} & \textbf{Identity \& Friction Obligations} \\ \midrule
1  & Personal diaries, hobby groups & Full pseudonymity; no legal-identity linkage.  Content governed only by ordinary community rules. \\[4pt]
2 & Niche influencers, local news pages & Platform-held verification of a government identity. Cooling-off window for posts; tamper-proof audit log retained. No public disclosure of real-world identity. \\[4pt]
3 & National media brands, celebrities & Independent, ML-assisted fact-checking and provenance watermarking \emph{before} algorithmic amplification; searchable public archive of corrections. Non-compliance triggers down-ranking or removal. \\ \bottomrule
\end{tabularx}
\end{table}

We employ \emph{friction} -- any deliberate cost or delay imposed on posting or sharing -- as a design principle. This has been shown to reduce misinformation and abusive speech by prompting deliberation. Empirical studies of social media and respective industry roll-outs of ``read-before-retweet'' or ``reconsider reply'' prompts demonstratively cut harmful interactions, meanwhile friction was shown to significantly improve the average quality of posts~\cite{hatmaker2020readprompt,vincent2020readprompt,jahn2023frictioninterventionscurbspread}. A tiered anonymity regime institutionalizes friction proportionally: identity verification and fact-checking occur only when content exceeds influence thresholds, preserving low-stakes spontaneity while dampening high-stakes manipulation.

We ground our proposal in using empirical evidence from Reddit's community moderation approach. Volunteer moderators already converge on proportional governance: as subreddit traffic grows, moderators introduce karma\footnote{In Reddit's context, this denotes an aggregate reputation metric equal to the net difference between positive and negative votes that a user's submissions and comments receive. It thus functions as a quantifiable proxy for community trust and is frequently employed as an eligibility threshold for posting or moderation privileges} minimums, manual reviews, or identity checks before posts appear~\cite{Kiene_2016,10.1145/3686920,Fiesler_Jiang_McCann_Frye_Brubaker_2018}. This means baseline pseudonymity for ordinary users is largely preserved despite stricter rules for influential posts. These organic practices demonstrate operational feasibility and indicate latent demand for tiered accountability that transcends any single platform architecture.

We further show that there is a viable regulatory pathway to achieving our proposal. The European Union's Digital Services Act already requires marketplaces to verify business users and offers a blueprint for identity-linked content duties~\cite{dsa}. Under the UK Online Safety Act, “Category 1” services must give adult users the ability to filter interactions with unverified or anonymous accounts and are required to make identity-verification mechanisms available, without imposing mandatory verification on all users ~\cite{UKOnlineSafetyAct2023}. By drawing on these precedents, legislators can embed tiered anonymity into safety models and ranking systems without prohibiting pseudonymity outright.

\paragraph{Contributions} This paper makes two main contributions:
\begin{enumerate}
    \item We introduce a formal model that maps user reach to escalating identity and verification duties, capturing both follower-heavy and suddenly viral accounts. This is supplemented by empirical evidence from a longitudinal Reddit case study that shows that proportional identity governance emerges endogenously in large online communities.
    \item We chart a jurisdiction-spanning regulatory pathway that leverages existing DSA and Online Safety Act provisions to operationalize the model.
\end{enumerate}

By calibrating identity obligations to influence, tiered anonymity restores proportionate friction to digital speech, aligns platform incentives with democratic values, and closes the accountability gap that AI-augmented misinformation eagerly exploits.

\section{Background on Deep Fakes and Mass Misinformation}\label{sec:background}

Deepfakes are realistic-looking but otherwise fabricated pieces of media usually used with malicious intent. They are typically produced by generative or diffusion ML models, and can be adapted to, for example, a target individual's voice and physical likeness, enabling malicious actors to create convincing impersonations of real people~\cite{perov2021deepfacelabintegratedflexibleextensible}. It is widely accepted that they pose a serious threat to the integrity of online information~\cite{Romero_Moreno01092024}. In practical terms, anyone with a moderately powerful computer or even a smartphone app can now superimpose someone's face onto another video or clone a public figure's voice with startling realism -- barriers to creating believable fake audio-visual content have all but evaporate~\cite{cooke2024crossing}. Critically, these forgeries are often difficult for humans to detect -- an experimental study found that people can no longer reliably distinguish authentic images or recordings from AI-generated fakes~\cite{cooke2025goodcointosshuman}. Consequentially, deepfakes erode trust in all media evidence, blurring the line between reality and fabrication. Legal scholars have warned of a resulting ``liar’s dividend'', wherein the mere existence of such convincing fakes allows bad actors to dismiss real inconvenient evidence as fake news~\cite{Chesney_Citron_2019}.

Compounding this challenge, large language models (LLMs) have drastically amplified the scale of text-based misinformation. LLMs like OpenAI's GPT-4 can generate polished persuasive text at a velocity and volume impossible for human propagandists to match. Traditionally, orchestrated disinformation required teams of trolls or paid writers to produce and disseminate false narratives~\cite{cartwright_2019}. Now, a single operative with an AI tool could mass-produce propaganda in numerous languages, tailored to different audiences, with a click of a button. Indeed, recent field research demonstrated that LLM-generated arguments on Reddit's \texttt{/r/ChangeMyView} were three to six times more persuasive than human posts~\cite{zurich2025ai_persuasion_experiment}. Similar findings were reported by~\citet{schoenegger2025largelanguagemodelspersuasive} adding that readers would be unlikely suspect AI authorship, making human-led detection and reporting difficult.

The \texttt{/r/ChangeMyView} field research further supports the effectiveness of a hypothetical ``AI-powered botnet'' that could ``seamlessly blend into online communities'' and manipulate public discourse~\cite{zurich2025ai_persuasion_experiment,Yang_2024}. Such setups could empower bad actors to overwhelm fact-checkers and exploit cognitive biases, accelerating misinformation dissemination on an unprecedented scale. Related to the above,~\citet{newsguard2025ai_tracking_center} recently identified 1\,200 unreliable websites that ran largely AI-composed news content in multiple languages, all with minimal human oversight. This new breed of automated disinformation outlet can churn out hundreds of articles a day. The LLM-produced text is grammatically correct, contextually relevant, and often superficially credible, easily fooling the average reader. Misinformation experts have accordingly dubbed generative AI ``the next great misinformation superspreader'', pointing out that AI systems empower almost anyone -- from hostile governments to conspiracy theorists -- to scale up deceptive content production dramatically~\cite{verma2023misinformation_superspreader}.

\section{Friction, Identity, and Accountability}

The starting point for any meaningful reform of online anonymity must confront a central tension in liberal democracies: the commitment to free expression versus the need to mitigate its weaponization. Classic accounts of speech rights, from John Stuart Mill to modern First Amendment jurisprudence, treat expression as a public good -- presumptively beneficial and self-regulating~\cite{campbell2017natural,mill1859liberty,jacobson2000mill}. Yet in algorithmically mediated environments, where LLMs can generate plausible falsehoods at scale and deepfakes blur the boundary between reality and fabrication, virality becomes decoupled from both truth and reputation, and the foundational assumptions underpinning these traditions lose coherence.

Friction -- in the form of verification, moderation, or traceability -- is often framed as a threat to the open internet~\cite{jahn2023frictioninterventionscurbspread}. That said, friction is a \textit{democratic design feature}~\cite{weidmann2019internet}. In physical communities, social friction arises from reputational consequences, geographic co-presence, and mutual visibility. One is less likely to spread inflammatory falsehoods in a town hall than online, not because one is more moral, but because the social costs are real and immediate. Digital platforms, in contrast, systematically dissolve these frictions. Recommender systems prioritize engagement, not deliberation; speed trumps reflection; and pseudonymity attenuates accountability~\cite{han2017revisiting}.

This breakdown of reputational checks facilitates what some scholars call ``context collapse'' -- the dislocation of speech from relational context~\cite{freelon2012talking,john2013sharing}. A post from a user with ten followers may be algorithmically amplified to ten million others without any change in content quality, intent, or reliability. The current legal system continues to treat users with ten followers as functionally identical to one with ten million followers. This is the core problem: the law protects anonymity \textit{symmetrically}, while platforms distribute influence \textit{asymmetrically}~\cite{Karanicolas2020Tackling,Skopek2016Anonymity}.

We argue that identity obligations must scale with content reach , particularly to mitigate the systemic risks posed by synthetic media as described in~\Cref{sec:background}. This is not a blanket call for real-name policies, which have been rightly criticized for silencing vulnerable speakers~\cite{cho2012empirical,macaulay2016queen}. Instead, it is a call for \textit{proportional identity calibration}~\cite{Cava2022Proportional}, wherein pseudonymity is preserved for low-reach users, while higher-tier actors must submit to private identity verification and, ultimately, to structured content review~\cite{Hiroyuki2016Pseudonyms}. This approach mirrors how democratic institutions already manage power: with increasing transparency and accountability as influence grows~\cite{Filgueiras2016Transparency}.

Our Reddit case study, described in~\Cref{sec:reddit_case_study}, illustrates this principle in practice. As communities expand in size and influence, moderation architectures evolve from permissive to hierarchical: identity checks, posting restrictions, and content approvals become the norm. These organically emergent structures reflect a collective intuition: that scale demands scrutiny and visibility must be earned.


\section{Reddit Case Study in Community Moderation at Scale}\label{sec:reddit_case_study}
Reddit offers a 20-year natural experiment in large-scale, bottom-up governance. More than 100\,000 active communities (subreddits) are overseen by roughly 60\,000 volunteer moderators who outnumber the platform's $\sim$400 paid administrators by two orders of magnitude~\cite{Strandell_2025,moderation_singh}. In the first half of 2024 alone, users generated 5.33 billion pieces of content; moderators and admins removed 3.1\% of it -- half by volunteers, 71\% of whose actions were automated by tools such as \texttt{AutoModerator}~\cite{Reddit_transparency_2024}. Unpaid labor on this scale has been valued at \${}3.4 million per year~\cite{li2022measuringmonetaryvalueonline}.

\paragraph{Multi-layer Moderation} Governance operates on three nested layers: (i) site-wide rules enforced by a small admin team, (ii) subreddit-specific rules defined and enforced by volunteer moderators, and (iii) crowd signals (voting, reporting) supplied by ordinary users.  Empirical analyses show that popular subreddits add \emph{more} and \emph{stricter} rules as audience size grows, often introducing karma thresholds, URL whitelists, or manual approval queues~\cite{Kiene_2016,Fiesler_Jiang_McCann_Frye_Brubaker_2018}. High-visibility communities even demand identity proofs: \texttt{r/BlackPeopleTwitter}, for example, required photographic skin-tone verification to curb impersonation~\cite{10.1145/3686920}. These organically emerging ``tiered'' signals parallel our proposed reach-based anonymity model.

\paragraph{Adaptive Structure} Reddit's structure evolves with scale and external pressure. In 2015, subreddits controlling much of Reddit's front page shut down (``AMAgeddon'') to protest inadequate mod tooling, prompting the company to invest in logs, modmail, and automated filters~\cite{Verge_2015}. In 2023 more than 7\,000 subreddits went private to oppose new API fees, again demonstrating the collective leverage of volunteer governance~\cite{Peters_Porter_2023}. Despite these confrontations, the core design -- local autonomy constrained by platform-level minima -- has remained intact and resilient.

\paragraph{Scale causes Friction}  Quantitative work finds a positive correlation between subreddit size and the likelihood of (i) entry gates (minimum account age/karma)~\cite{Fiesler_Jiang_McCann_Frye_Brubaker_2018}, (ii) pre-publication queues~\cite{Fiesler_Jiang_McCann_Frye_Brubaker_2018,bhattacharya2024unveilingscalinglawsregulatory}, and (iii) ex post identity checks~\cite{10.1145/3686920}. For example, nearly all large subreddits quietly enforce minimum karma or account-age requirements: smaller communities may require 50 karma, whereas top-tier subreddits often demand well over 1,000 karma points~\cite{10.1145/3375197}. In some high-visibility subreddits, moderators verify a user’s real identity privately for credibility (e.g.\ the \texttt{/r/IAmA} forum requires proof of identity for hosts) while allowing the user to remain publicly under a pseudonym. In other words, moderators intuitively impose \emph{proportional friction}: low-reach users post freely; higher-reach content encounters verification or review. Reddit thus supplies real-world evidence that tiered anonymity is operationally feasible and socially accepted when the costs of influence are borne chiefly by those who wield it.

\interfootnotelinepenalty=10000
\begin{table}[t]
\centering
\caption{Identity and moderation norms on major social-media platforms as of May 2025. ``Tiered'' denotes any mechanism in which obligations or scrutiny escalate with audience size or monetization status\protect\footnotemark.}\label{table:platform_comparison}
\begin{tabularx}{\textwidth}{@{}l c X@{}}
\toprule
\textbf{Platform} & \textbf{Community Moderation} & \textbf{Tier-like content checks} \\ \midrule
\multirow{3}{*}{Reddit} & \multirow{3}{*}{Volunteer moderators} 
  & Karma / account-age gates \\
 & & \texttt{AutoModerator} keyword filters \\ 
 & & Stricter rules as subreddit size grows \\ \midrule
\multirow{2}{*}{Facebook} & \multirow{2}{*}{No} 
  & Centralised review by staff and contractors \\ 
 & & No escalation tied to reach \\ \midrule
\multirow{2}{*}{Instagram} & \multirow{2}{*}{No} 
  & Feature gates at $\sim$10k followers (links, product tags)\\ 
 & & Content demotion or removal on policy breach \\ \midrule
\multirow{2}{*}{X (Twitter)} & \multirow{2}{*}{Community Notes} & No systematic reach-based review \\
& & Enforcement tied to policy breaches\\ \midrule
\multirow{2}{*}{TikTok} & \multirow{2}{*}{No} 
  & Increased human review for high-follower creators  \\
 & & Scaled ML enforcement for long-tail users \\ \midrule
\multirow{3}{*}{YouTube} & \multirow{3}{*}{No} 
  & Automated checks for new channels \\
 & & Manual review for Partner-Program content ($\geq1$k subs) \\ 
 & & Additional scrutiny for $100$k+ channels \\ \bottomrule
\end{tabularx}
\end{table}\footnotetext{Note: Both Facebook and Instagram have limited community moderation in the form of user reports, content-flagging tools, and basic guideline enforcement. However, this does not equate to the Reddit model, where moderation is structurally decentralized: individual subreddits are governed by volunteer moderators with delegated authority to set rules, curate discourse, and shape community norms. Facebook and Instagram lack this layered, self-governing architecture; their ``community moderation'' is reactive, not constitutive.}

\paragraph{Contrast with Centralized Platforms}  Competing platforms provide no comparable venue for community-level rule-making.  Facebook real-name enforcement, X's paid ``blue check'', and YouTube's purely algorithmic filters all exemplify top-down moderation with minimal local discretion or tiering. Comparative studies confirm that Reddit alone relies ``more or less on self-moderation by volunteers'', producing a distinctive, multi-layer oversight regime~\cite{Beijbom_2022}. We summarize our findings regarding identity and moderation norms in all current major social media platforms in~\Cref{table:platform_comparison}. YouTube provides the closest analogue to our tiered system: ``new'' channels face automated checks, Partner-Program creators add identity and monetization audits, and six-figure-subscriber channels receive further manual review and provenance badges.

\paragraph{Take-Away}  Reddit's layered system demonstrates that \emph{identity obligations can scale with reach} without eroding baseline pseudonymity. The empirical pattern -- stricter gates as audiences expand -- mirrors the normative logic of our three-tier framework and supplies a practicable blueprint for regulatory codification on platforms that lack subreddit-style boundaries.

\section{Proposed High-level Technical Implementation}\label{sec:high_level_implementation}

Most large platforms already track granular engagement telemetry (followers, impressions, reshares, watch-time). We propose aggregating these signals into a rolling \emph{reach score} mapped to tier thresholds proposed in~\Cref{table:tiers_summary}. As engagement models differ by platform, each platform must calibrate threshold values to its context. For fairness, \textit{reach} can be defined in relative terms (e.g.\ top percentile of active users) to achieve consistency across services. A post that spikes above a threshold would trigger retroactive elevation of the account’s tier. This design ensures that one-off virality is handled without allowing bad actors to repeatedly skirt the limits. Refining the exact reach formula and threshold calibration is left for future work; it will be crucial to balance curbing abuse with not unduly deterring mid-tier creators from legitimate growth.

Platforms should complement hard metrics with contextual triggers such as monetization enrollment or activation of business tools. Precedent exists with Instagram withholding external–link ``Swipe-Up'' stories until an account reaches \(\sim\!10\,000\) followers or holds a business profile, effectively coupling functionality to influence~\cite{Walker_2024}. A similar gating mechanism can enforce tier promotion automatically while minimizing false positives.

\paragraph{User-facing Controls} The UK {Online Safety Act 2023} obliges Category 1 services to provide adults with filters that exclude non-verified users~\cite{UKOnlineSafetyAct2023}. A tiered system can generalize this idea: clients expose a preference pane that lets users down-rank or hide Tier 1 content, surface fact-check banners for Tier 3 posts, or receive warnings when resharing material from unverified sources. Such controls translate legal duties into actionable UX.

\paragraph{Tier Details} We provide some illustrative guidelines for our proposed tiers: \begin{enumerate}
    \item[\emph{Tier 1}] No additional obligations: posts remain subject only to baseline community rules.
    \item[\emph{Tier 2}] Accounts must complete \emph{private} identity verification and comply with advertising-law disclosure.  The US\ FTC's \emph{Endorsement Guides} require influencers to reveal any ``material connection'' with brands in a manner that is ``clear and conspicuous''~\cite{ftc_influencers_2019}. Automated classifiers can flag suspected undisclosed ads for moderator review.
    \item[\emph{Tier 3}] High-reach accounts are treated as de-facto publishers.  Posts containing political, health, or financial claims are routed -- before wide distribution -- to an external fact-checking queue.  Empirical surveys by UNESCO show that 62\% of digital creators do not verify information before sharing, underscoring the need for mandated review~\cite{UNESCO_digital_creators_2024}. Provenance watermarks and a public correction log close the feedback loop; serious or repeated violations trigger algorithmic down-ranking or suspension.
\end{enumerate}

\paragraph{Progressive Friction} Existing platform tooling provides technical backing to ensure the necessary friction is applied:  
\begin{itemize}
  \item \emph{Rate-limited publishing queues} that lengthen with tier: seconds for Tier 1, minutes for Tier 2 (cool-off), hours for Tier 3 pending fact-check.  
  \item \emph{Priority triage} of user reports: complaints about Tier 3 content land at the top of moderator dashboards.  
  \item \emph{Automated provenance signals} (e.g.\ C2PA hashes~\cite{C2PA_2023}) injected at upload time for Tier 3 media, enabling rapid debunking should manipulations surface.  
\end{itemize}

These mechanisms impose costs proportionate to communicative power while leaving ordinary pseudonymous speech largely untouched, thereby operationalizing the normative principle that \emph{influence entails accountability}.

\section{Current Legal Precedents and Regulatory Infrastructure}

\subsection{European Union:  Evolving From KYBC Toward Influence-Based Identity Controls}

The European Union provides the strongest foundation for codifying tiered identity obligations. The Digital Services Act (DSA) already introduces structural mechanisms that can be repurposed to support a reach-based verification regime. Article 30's \textit{Know Your Business Customer} (KYBC) requirement, which mandates identity verification for commercial users, represents a conceptual shift: platform functionality is increasingly conditioned on user transparency~\cite{eu_dsa_article30}.

More significantly, Articles 34 and 35 impose systemic risk obligations on Very Large Online Platforms (VLOPs) -- defined by a monthly audience threshold -- explicitly linking platform reach to responsibility~\cite{eu_dsa_articles34_35}. These risk obligations are particularly salient in the age of generative AI, again linking back to the discussion in~\Cref{sec:background}. This capacity decouples virality from veracity, amplifying the urgency of proportional identity and transparency duties for high-reach users. This sets a critical precedent: the broader a user or platform's influence, the greater the required diligence. Article 10 further enables identity disclosure in response to illegality, reinforcing a principle of proportionality that mirrors the core logic of tiered anonymity.~\cite{EU2022DSA_Art10}.

Moreover, complementary frameworks like the AI Act and formerly proposed AI Liability Directive further strengthen this trajectory. By requiring labeling of synthetic media and audit trails for AI systems, the EU is already enforcing traceability in high-risk communicative environments~\cite{eu_ai_act_2024, eu_ai_liability_directive_2022}. Our proposed tiered framework fits squarely within this expanding digital \textit{acquis}. These instruments, taken together, suggest that scalable identity obligations based on content reach are not only compatible with EU law -- they are its logical extension.

\subsection{United Kingdom: The Online Safety Act and Voluntary Verification}

In contrast, the UK's Online Safety Act 2023~\cite{UKOnlineSafetyAct2023} establishes a statutory duty of care on digital platforms, particularly those classified as Category 1 services -- platforms with significant reach and functionality. Under the accompanying Categorization of Regulated Services (Threshold Conditions) Regulations 2024~\cite{UKThresholdRegulations2025}, these platforms are required to offer adult users the option to verify their identity and to provide tools enabling content filtering based on verification status. This framework embeds optional identity verification with a reputational signalling layer, while preserving the right to anonymity, laying the conceptual groundwork for our proposed tiered identity regime.

However, this identity framework remains voluntary and reputational rather than mandatory and enforceable. Users may choose to verify themselves, and others may opt to filter content accordingly; but no binding obligations are imposed on high-reach users who remain anonymous. Existing legal tools expose both the possibilities and the limits of the current system. Norwich Pharmacal Orders can compel disclosure only when platforms already hold identity data, a weak point given that many accounts have never been verified. \cite{norwich_pharmacal_order} Similarly, the UK–US CLOUD Act facilitates law-enforcement access to stored data but is not an identity-disclosure regime and does not ensure that platforms possess verified identities.{\cite{uk_us_cloud_act_agreement} These mechanisms show that anonymity is constrained in principle yet remain inadequate for ensuring accountability among high-reach anonymous users.

The current UK regulatory landscape lacks a proactive mechanism linking user influence -- measured by visibility, engagement, or monetization -- to identity obligations. We argue that this omission is increasingly untenable in an era of algorithmic virality where individuals can rapidly attain significant reach with little to no accountability.

A logical evolution of the Online Safety Act would be to mandate identity verification for users who exceed a defined influence threshold, as discussed in~\Cref{sec:high_level_implementation}. Such a reform would convert identity verification from a reputational indicator into a mechanism of enforceable accountability. By embedding this obligation within the existing statutory framework, the UK could pioneer a \textit{rights-preserving} yet \textit{responsibility-tiered} \textit{model} of online governance -- one that maintains anonymity for everyday users while ensuring that high-reach actors meet proportionate standards of transparency and legal traceability.

\subsection{United States: First-Amendment Boundaries and Conditional Immunity}

The United States presents the most challenging jurisdiction for any form of compelled identity regulation due to robust First Amendment protections and the shield of Section 230\footnote{Section 230 provides that online platforms are not treated as the ``publisher'' of user content, effectively immunizing them from most civil liability for third-party posts} of the Communications Decency Act~\cite{section230}. American courts have repeatedly upheld the right to anonymous speech, particularly in digital spaces. Landmark cases such as \textit{Doe v. Cahill} and \textit{Dendrite Int'l, Inc. v. Doe No. 3}~\cite{doevcahill2005} require plaintiffs seeking to unmask anonymous users to meet stringent standards, such as presenting a \textit{prima facie} case of harm and passing a balancing test that weighs the speaker's right to anonymity. 

Despite this, momentum is growing at the federal level toward rethinking the blanket nature of Section 230 immunity. Legislative proposals, including bipartisan efforts, have increasingly considered conditioning immunity on a platform's compliance with transparency and good-faith content moderation practices~\cite{sinnreich2022, dickinson2023}. Crucially, however, the constitutional limits on compelled identity disclosure derive from the First Amendment, not from Section 230. Section 230 protects platforms from liability for third-party content; it does not shield them from disclosure obligations or other regulatory requirements, which are instead constrained by First Amendment doctrine. Recent legislative steps also target AI-driven harms directly -- for example, the 2025 TAKE IT DOWN Act criminalizes the distribution of non-consensual intimate imagery (including AI-generated deepfakes) and mandates swift removal of such content~\cite{USCongress_S146_119th_2025}. Rather than mandating universal real-name usage (which would invite constitutional challenges), these proposals suggest a path for indirect, incentive-based regulation that respects constitutional limits while introducing mechanisms of accountability.

In this context, our proposed tiered identity approach offers a legally viable and technically feasible approach. Platforms could retain full Section 230 protections only if they implement a reach-based anonymity tier system. Basing the framework on \textit{influence}, or other platform-side metrics such as engagement telemetry and monetization enrolment, allows the system to remain content-neutral, which is crucial for surviving constitutional scrutiny~\cite{goodman2019,cramer2020}

Furthermore, this model dovetails with user-choice provisions already emerging in US and UK law. US platforms could offer analogous controls to the UK's filtering out of unverified users~\cite{UKOnlineSafetyAct2023} -- such as the ability to down-rank Tier 1 content or flag Tier 3 posts with fact-check banners.

In sum,  we believe that basing our tiered framework on influence rather than identity per se provides a constitutionally sound middle ground. It operationalizes the principle that ``influence entails accountability'', not by restricting speech, but by assigning procedural obligations where amplification is algorithmically enabled~\cite{park2024}. 

\section{Piercing Anonymity and Legal Thresholds}

While previous sections have outlined the legal mechanisms available to unmask anonymous actors, this section turns from retrospective tools to the conceptual and operational implications of prospective identity collection -- that is, requiring platforms to obtain verifiable identity data from users before harms occur, based on the scale of their content reach.

Legal regimes in the EU, UK, and US all permit ex post identity disclosure in narrowly defined circumstances. However, these mechanisms often prove too slow or reactive for mitigating fast-moving misinformation. Courts and regulators typically intervene only after content has already spread and caused damage at which point the harm is often irreversible~\cite{rogal2013,froomkin1999}. This is particularly true in cases involving LLM-generated misinformation or synthetic deepfake videos, which can go viral in minutes and be nearly impossible to recall once distributed. As generative models become more accessible and realistic, the cost of delayed identity resolution rises exponentially. 

The tiered anonymity model proposed here shifts this paradigm. For Tier 2 and Tier 3 users -- those with moderate to large followings -- platforms would be required to collect and securely store legal identity information \textit{in advance}, subject to minimal access protocols and stringent privacy protections. This would allow for swift disclosure upon valid legal request while protecting pseudonymity in everyday use. The goal is not to reduce anonymity universally, but to \textit{contextualize it based on communicative power}~\cite{nagel2015, candeub2017}.

Crucially, this shift does not necessitate the rewriting of existing legal thresholds for unmasking identities. Rather, it enhances procedural efficiency and evidentiary readiness when those thresholds are met. For example, a court order that might normally take weeks to execute due to jurisdictional barriers and technical resistance could be processed swiftly if the platform has already verified identity and established a lawful disclosure protocol~\cite{cho2012empirical}.

To preserve civil liberties, identity databases must be governed by robust safeguards. These include: \begin{itemize}
    \item \textbf{End-to-end encryption} for stored identity data
    \item \textbf{Access logging} to track who requests and receives information
    \item \textbf{Data minimization} (collecting only what is necessary)
    \item \textbf{Retention limits} with periodic review and deletion
    \item \textbf{Cross-border legal harmonization}, particularly through mutual legal assistance treaties, and agreements like the CLOUD Act~\cite{uk_us_cloud_act_agreement}
\end{itemize}

This approach reframes identity not as a binary attribute, but as a \textit{regulated credential}: conditionally disclosed, proportionately applied, and safeguarded by due process. As such, it avoids the pitfalls of approaches such as South Korea's real name policy (detailed in the following section) while addressing the increasing costs of untraceable amplification~\cite{park2012}. Ultimately, prospective identity collection enables responsiveness without repression: a legal architecture suited for the velocity and asymmetry of the contemporary information ecosystem.

\pagebreak\section{The Global Momentum for Conditional Pseudonymity}

The international policy environment is already increasingly converging around the idea that identity obligations should scale with user influence. Among the first state efforts to discipline online anonymity was South Korea’s real-name verification regime, a sweeping identity-disclosure mandate intended to eliminate anonymous participation across major platforms.
That approach proved both legally unsustainable and practically ineffective. The Korean Constitutional Court invalidated the policy for violating freedom of expression, and subsequent research showed it failed to reduce online harms in any measurable way~\cite{park2012,leitner2009}. The lesson was clear: blanket identity mandates are blunt instruments that overreach without precision.

Since then, regulatory energy has shifted toward more granular, influence-sensitive models. In India, the 2023 Draft Digital India Bill introduces a risk-based classification framework for digital intermediaries, suggesting a shift toward more nuanced regulatory obligations based on the type and scale of service; without explicitly extending these obligations to individual users or calibrating them to user influence~\cite{india2023digital}. Australia's eSafety Commissioner has advanced similar proposals, calling for the traceability of high impact accounts, particularly those linked to harmful or AI-generated content~\cite{esafety2024}. Meanwhile, the European Commission has initiated consultations on ``influence transparency'', exploring how verification requirements might apply to accounts disseminating politically sensitive or synthetic media~\cite{dsa,eprs2021deepfakes}.

Platform ecosystems increasingly reflect this logic, though in a fragmented manner. Meta's Verified program, X's (formerly Twitter) ``blue check'' system, and YouTube's monetization criteria all condition algorithmic reach, visibility, and revenue on voluntary identity disclosure or engagement thresholds~\cite{nehra2024,xiao2023verification,dunna2022paying}. These systems reinforce a de facto hierarchy -- creators with broader audiences receive preferential treatment while also facing greater scrutiny -- forming an implicit structure of tiered governance. However, these frameworks often lack transparency, consistency, and regulatory oversight~\cite{caplan2020tiered}.

Taken together, these developments suggest the emergence of a normative shift: pseudonymity remains appropriate for ordinary users, but must give way to verification and procedural safeguards -- identity linkage, content moderation, or algorithmic throttling -- once a user's reach crosses a defined threshold. We term this evolving model \textit{conditional pseudonymity}: a regulatory philosophy that preserves privacy for the many while introducing graduated accountability for the influential.

We build on this global momentum and consider our proposed tiered framework a formalization of trends already unfolding across jurisdictions and platforms, rather than a wholly new system. By codifying conditional pseudonymity, we provide a principled, scalable model rooted in proportionality and procedural fairness. It aligns regulatory tools with the actual distribution of digital power -- preserving pseudonymity where appropriate, qualifying it where necessary, and ultimately ensuring that privacy and accountability evolve in tandem in the algorithmic public sphere.

\section{Enforcement Challenges and Safeguards}

Implementing a tiered anonymity framework in a globally interconnected internet ecosystem presents significant enforcement challenges -- large commercial platforms operating across jurisdictions may resist added verification burdens, and users might evade restrictions by migrating to less-regulated services or using technical workarounds (e.g.\ bots to artificially reset reach). While Reddit shows that moderation hierarchies can emerge organically, its reliance on volunteer governance is difficult to replicate on more commercial platforms like Meta, YouTube, or X. These platforms operate across multiple jurisdictions but often default to the legal norms of their home country -- typically the United States -- resulting in fragmented regulatory oversight~\cite{blochwehba2018platform}.

To scale tiered anonymity, enforcement must be institutional, driven by governments and platforms rather than individual users. This coordinated institutional response is essential: voluntary self-regulation by platforms alone has thus far failed to curb disinformation, prioritizing profits over safety -- ``fundamentally self-regulation by social media platforms has failed to achieve the promised public policy benefit of improving the quality of the information ecosystem''~\cite{Hoffmann2019MarketDisinformation}. Governments in regions such as the EU, UK, and US already exercise regulatory authority over platforms operating within their borders. This authority can be extended extraterritorially, as seen with the EU's General Data Protection Regulation (GDPR), which extends obligations beyond EU borders through data adequacy requirements and reputational enforcement mechanisms~\cite{pramesti2020extraterritoriality, azzi2018gdpr}.

In lower-regulation or infrastructure-poor jurisdictions, implementation could be supported by interoperable digital identity standards that align with GDPR principles of data minimization, purpose limitation, and secure storage. Public-private partnerships or open-source systems, such as the European Digital Identity (EUDI) Wallet or India's Aadhaar infrastructure (with appropriate safeguards), could provide privacy-preserving verification without broad data disclosure~\cite{larsen2023identitywallets, chowdhry2021aadhaar, wagner2021eudi}.
 
However, unilateral regulation risks being seen as digital imperialism, especially in the Global South~\cite{jin2013platformimperialism}.To address this legitimacy challenge, multilateral cooperation is essential. Institutions such as the United Nations Internet Governance Forum, the OECD, and regional organizations like the African Union and ASEAN can serve as venues for aligning policies and establishing shared norms~\cite{antonova2014digitaldivide, bhuiyan2014internetgovernance}. Soft-law instruments -- non-binding principles, technical standards, and voluntary codes of conduct -- can serve as transitional tools toward global harmonization~\cite{flew2018technologytrust, shcherbovich2017multistakeholder}.

Crucially, strong safeguards must be in place to prevent misuse of this system by authoritarian regimes. A tiered identity framework must be framed as rights-preserving: it does not eliminate anonymity, but conditions large-scale amplification on accountability, while protecting vulnerable users like whistleblowers and activists. Authoritarian governments might otherwise attempt to weaponize identity requirements to unmask and persecute dissidents. We note that China, for example, has required real-name registration for ``self-media'' accounts over 500,000 followers – a move that prompted some users to shut down their accounts for fear of surveillance~\cite{FreedomHouse_FOTN_China_2024}. To guard against such abuse, our proposal insists on robust due-process limits on identity disclosure (judicial oversight, high thresholds for requests, and transparent challenge mechanisms)~\cite{zarsky2013regulating}. Stored identity data should be encrypted and access logged, with strict minimization and retention policies. These measures, combined with independent oversight, ensure that \textit{prospective identity collection enables responsiveness without repression}. The goal is a system where accountability scales with influence, yet privacy and free expression remain protected for the vast majority of users.

Resistance will likely be inevitable, especially from the platforms' side. They may object to the complexity and cost of implementation or worry about user attrition if stricter identity rules push users to fringe platforms. Likewise, some users may attempt to evade the implementation (as per the example from China above) by migrating to less-regulated services. To address this, enforcement must be both staged and strategic. High-leverage jurisdictions like the EU, UK, and US can drive adoption by linking regulatory compliance to market access (e.g.\ conditioning app store listings or ad business on tiered verification~\cite{flew2018technologytrust}). As discussed a few paragraphs above, multilateral cooperation is equally critical to avoid a fragmented patchwork of rules.

Ultimately, tiered anonymity is not about censoring speech; it is about regulating amplification in an era of LLM-driven mass-misinformation and deepfake proliferation. The tiered anonymity system does not erode privacy; it recalibrates the power asymmetry between digital citizens and systemic actors, advancing a more just and resilient online ecosystem that balances freedom of expression with democratic accountability and scales privacy with, rather than against, responsibility.

\section{Conclusion}

We advance a single claim: \emph{governments should require social-media platforms to calibrate anonymity to communicative reach}. By analyzing the epistemic harms of deepfakes and LLM-assisted misinformation, we show that the traditional symmetry of online anonymity no longer maps onto the asymmetry of algorithmic amplification. Our three-tier framework operationalizes the principle that \emph{influence entails accountability}: Tier 1 preserves full pseudonymity, Tier 2 introduces private identity linkage, and Tier 3 imposes publisher-level duties of verification and provenance.

This model \textit{extends}, rather than upends, current legal trajectories -- for example, the EU’s DSA~\cite{dsa} and UK’s Online Safety Act~\cite{UKOnlineSafetyAct2023} already link greater reach to greater responsibility, and even in the US recent laws and proposals such as the TAKE IT DOWN Act~\cite{USCongress_S146_119th_2025} and Section 230~\cite{section230} are trending toward conditional accountability. Adopting tiered anonymity would re-introduce the social friction that recommender systems have eroded, dampening the incentive and impact of large-scale disinformation while sparing ordinary users onerous disclosure.

Future work can focus on refining threshold criteria and developing privacy-preserving credential solutions to further bolster the framework. We are confident that calibrating anonymity to influence offers a viable path to a healthier online ecosystem -- one that preserves the liberating \textit{dog-on-the-Internet}~\cite{steiner1993dog} anonymity for millions of users, yet ensures that those who command the attention of millions are no longer shielded by total anonymity at the expense of the public good. Each additional safeguard and iteration will bring us closer to an internet that balances free expression with accountability, and privacy with responsibility, in the age of AI-generated media.

\begin{ack}
David Khachaturov is supported by the University of Cambridge Harding Distinguished Postgraduate Scholars Programme.
\end{ack}

\bibliography{main}

\end{document}